# Alternative AI Philosophy: Daoism as Method for AI in Education

Qin Xie, University of Minnesota, USA

## Abstract

As artificial intelligence (AI) rapidly iterates and transforms teaching, learning, and knowledge production, philosophical reflection has become increasingly indispensable to educational debates still predominantly shaped by Western intellectual traditions. This article proposes Daoism as an alternative philosophical framework for reimagining AI in education. Through philosophical analysis and textual interpretation of classical Daoist sources in dialogue with contemporary scholarship on AI in education, it examines how the Daoist concepts of "Dao nature," "self-cultivation," and the "Zhenren" address fundamental questions concerning reality, the epistemic aims of education, and ethical action in the AI-mediated era. In doing so, the article diversifies the philosophical voices shaping inquiry into AI and education, enriching the field's conceptual resources and offering a genuinely pluralistic foundation for comparative philosophy of education in the AI era.

## Introduction

More than three decades ago, John McCarthy, a foundational figure in artificial intelligence, contended that "AI needs many ideas that have hitherto been studied only by philosophers" (McCarthy 1995, 2041). At that time, however, such philosophical engagement was not yet imperative, as AI systems were not required to operate autonomously within complex real-world environments. McCarthy argued that once artificial intelligence approaches human-level capability, it becomes necessary to endow programs and algorithmic models with philosophical frameworks, particularly epistemological ones. In light of subsequent technological advances, McCarthy's prescient observation now appears not only vindicated but increasingly urgent. On February 24, the Federal Reserve Bank of New York (2026) reported that in 2024 U.S. philosophy graduates had a lower unemployment rate than their computer science counterparts, 5.1% versus 7%, a finding that challenges prevailing assumptions about the employment prospects of humanities graduates relative to technical fields (Laumer et al. 2012). As computing power grows and AI democratizes technical knowledge, lowering barriers in domains such as coding and programming (Dessimoz and Thomas 2024), distinctly humanistic capacities become comparatively more valuable.

As AI-facilitated knowledge democratization continues, the demand for humanistic and philosophical inquiry into AI's integration into society grows more urgent. Faced with abundant thorny societal problems (e.g., when computer vision misidentifies a face, who bears the consequences? When a driverless car kills someone, who is responsible?), technical solutions alone cannot resolve them. Major AI labs have therefore increasingly hired philosophers and incorporated philosophical principles into the design and alignment of AI models. Anthropic, for example, has embedded Kantian ideas into Claude's AI constitution under philosopher Amanda Askell (Sutherland 2026), and philosopher Harvey Lederman has incorporated Wang Yangming's (王阳明) notion of "Genuine Knowledge" (真知) into Claude's design and training (Lederman 2022). Starting with Claude Haiku 4.5, the first generation to operationalize Wang's philosophical framework, all later iterations have demonstrated substantial progress and consistently attained top-tier performance on the AI agentic misalignment benchmark (Anthropic 2026).

The growing integration of philosophical expertise into AI development implicitly answers the fundamental question posed by AI pioneer Alan Turing: "Can a machine think?" Contemporary AI models readily surpass the Turing test, and numerous systems now outperform humans in certain domains (Grace et al. 2018). AI has therefore rapidly integrated into human society and touches the very site of cognitive cultivation: education. For instance, the Washington, D.C. public school system implemented IMPACT, an AI-powered tool designed to identify and remove ineffective teachers; Tennessee introduced the Value-Added Assessment System (TVAAS) to evaluate teacher performance; and the Houston Independent School District deployed EVAAS for the same purpose. Algorithmic management of teaching and learning, however, can fail to respect individual agency and produce harmful moral consequences (Rubel et al. 2020). The release of ChatGPT marked a further turning point, as generative AI profoundly reshapes the entire education ecosystem (Al-Zahrani 2024); AI is no longer merely a smart system or facilitation tool but has "become" a learning partner and research confederate, whether openly or covertly, participating in teaching, learning, and knowledge production (Costa et al. 2025).

Debate on the promises and perils of AI in education has heightened significantly since late 2022. While current scholarship focuses largely on applied dimensions, such as policy design, academic integrity, and best practices, these pragmatic concerns, though crucial, are only part of the picture (Baidoo-Anu and Owusu Ansah 2023). Any serious conversation on these issues inevitably returns to philosophical questions of being and reality: *What is truth, and what is real about AI in education? What are the epistemic aims of education in an AI-shaping world? How should we live in such a transformed epistemic environment?*

Philosophers and scholars have begun to address these questions, yet their accounts share a common architecture: they are mediated predominantly through Western philosophy, which treats epistemic agency as a property of discrete subjects, frames knowledge in representational and calculative terms, and locates the human-AI relation within a subject-object dualism. Ferrario et al. (2024) critique the presumed epistemic superiority of AI systems and propose a human-AI hybrid epistemic agent, yet this still presupposes the individualist model of agency it seeks to revise. Driggers and Boyles (2025) draw on Peirce's pragmatism to evaluate appropriate AI uses in education, but pragmatist criteria of warranted belief remain anchored in Western empiricist accounts of inquiry. Cope et al. (2021) argue that, given the limits of transposing human meanings into numbers, AI can only serve as a "cognitive prosthesis" rather than a genuine educator, a critique that assumes the representationalist gap between qualitative meaning and quantitative form. Cox (2024) employs the metaphors of "Makers," "Managers," and "Inforgs" to reconcile educational goals with new conceptions of epistemic agency, but these metaphors compete within, rather than question, a broadly liberal-humanist picture of the knower. Because Western epistemology underwrites both the design imaginary of AI systems and the evaluative criteria applied to them, its limits become the limits of the field's collective imagination.

Scholars such as Sylvia Wynter (2003) and Lisa Lowe (Lowe 1994; Lowe and Manjapra 2019) have shown that this Western lens is not culturally neutral but entangled with imperial and colonial frameworks of the human, rationality, and knowledge. If the concepts through which AI in education is understood are themselves products of a particular tradition, a genuinely comparative approach is needed, not to add "Eastern content" to an unchanged framework, but to interrogate the framework itself. Wang Yangming's philosophy has already contributed meaningfully to addressing AI agentic misalignment, demonstrating that non-Western traditions can serve as generative sites of theory rather than peripheral sources of content. This calls for deeper engagement with non-Western wisdom traditions to pluralize knowledge production (Chen 2021), while avoiding the pitfalls of Orientalism (Said 1995). Among such traditions, Daoism is an especially promising comparative partner: where the Western accounts above presuppose subject-object dualism and calculative rationality, classical Daoism explicitly problematizes both, offering an alternative ontology of knowing, acting, and cultivating that has already informed educational thought (Kirkland 2004).

This article proceeds through conceptual analysis and comparative interpretation. In what follows, I first provide a brief introduction to Daoism and clarify why it is distinct from Confucianism, the other major Chinese philosophical tradition in the AI-shaping education context. I then elaborate the Daoist concept of "Dao nature" (道性) and explore its potential to address fundamental questions of truth and the epistemic aims of AI in education. In the subsequent section, I examine the Daoist conception of "self-cultivation" in order to answer how we ought to act in relation to AI in education. Having established these "what" and "how" dimensions, I propose one Daoist practitioner role, the "Zhenren" (真人), as a new alternative myth for AI in education, specifically for reimagining learning.

## "Dao Nature (道性)" as Epistemic Aim of AI in Education

### *What Is Daoism, and Why Not Confucianism?*

For more than two millennia, Daoism (also spelled "Taoism") has been integral to Chinese intellectual and everyday life; its central concept is widely regarded as the most fundamental, profound, and notoriously difficult to define in the history of Chinese philosophy (Wieger 1976). Daoism is traditionally traced to the late Spring and Autumn period (春秋时期, sixth century BCE), an era of political turbulence that saw a flourishing of ancient philosophical schools, later known as the "Hundred Schools of Thought" (百家), each offering competing visions of order. Daoism, with Laozi (老子) as its founding figure, stands as one of these schools (Zhao 1986; Li 2009; Hu 2005).

The philosophical ideas of Laozi and his followers gradually coalesced into a distinct school of thought known as Daojia (道家), or Philosophical Daoism. This tradition is grounded in a canon of foundational texts, such as the *Zhuangzi (庄子), Neiye (内业), Daodejing (道德经)*, and *Huainanzi (淮南子)*, each offering a practical and contemplative way of life (Wieger 1976). Over time, these philosophical teachings were absorbed into and transformed by an organized religious tradition known as Daojiao (道教), or Religious Daoism, which encompasses a complex constellation of institutional structures, folk beliefs, rituals, spiritual practices, and sacred texts including the *Taipingjing (太平经)* (Robinet 1997). Accordingly, Daoism is broadly understood as comprising two interrelated but distinct dimensions: Philosophical Daoism (道家) and Religious Daoism (道教).

However, scholars have increasingly challenged this philosophical/religious dichotomy, arguing that the two dimensions have been historically inseparable. Early Daoist texts, for instance, are replete with references to spiritual and meditative practices, suggesting that the boundary between philosophy and religion was never as clear-cut as the categorization implies (Coutinho 2013). Kirkland (2004) offers a particularly compelling reframing: Daoism is best understood not as a fixed doctrine to be classified but as what Daoists do, a living, practice-centered tradition defined by "the accumulated facts of Daoist practice," embodied by those whose ideas, values, and practices are expressed in the vast collection of texts known as the *Daozang (道藏)* and its continuations (2004, 13). Daoists absorbed the texts preserved in the *Daozang*, internalizing their teachings as practical guidance for living. Among all known Daoist texts, the *Daodejing*, the foundational treatise on the Principle and its Action, stands as the central doctrine of Daoism.

This canonical text expounds on *Dao (道)* and *De (德)*, commonly rendered as the "Way" and its "Virtue" (Laozi 2019). The essence of the Dao is constituted by two immanent and complementary properties: *Yin (阴)* and *Yang (阳)*. At the heart of the *Daodejing* is the concept of *Wuwei (无为)*, non-action or the absence of unnatural, forced intervention (Huang 1991). Dao itself takes no deliberate action, yet sustains and supports all things in their natural states, a naturalistic orientation that stands in direct opposition to artificial regulation, social engineering, and ceremonial convention (Zhan and Xie 2009).

Daoism thus advocates a spontaneous and harmonious way of life aligned with the natural order of the Dao. This vision contrasts sharply with Confucianism, the other foundational philosophical tradition in China, which emphasizes social hierarchy, effortful moral cultivation, and structured ethical conduct (D'Ambrosio 2020). Central to Confucian thought is the preservation of social order through the faithful fulfillment of role-based obligations and ritualized propriety. The cultivation of personal virtues, benevolence (仁), righteousness (义), propriety (礼), wisdom (智), and fidelity (信), ultimately serves to uphold and reproduce the prevailing patriarchal power structure (Herr 2014). In brief, Daoism emphasizes the natural flow of existence, while Confucianism highlights the maintenance of social order. Narrowing to the educational context, the two traditions espouse fundamentally different aims of education.

In Confucian thought, education is explicitly oriented toward governance and social utility. In the *Analects (Lunyu, 论语)*, one of Confucianism's foundational texts, Zixia states: "Given an excess of virtue in the conduct of one's official duties, one should study. Given an excess of virtue in one's studies, one should hold official duties" (仕而优则学，学而优则仕) (Ames and Rosemont 1999, chap. 19). This passage encapsulates the Confucian ideal of entering officialdom through learning, which became the defining expression of the "reading books to become an official" (读书做官) mindset in traditional Chinese society (Gardner 2007). Education, in this framework, serves governance and, by extension, the preservation of social structure and hierarchical order. Daoism, by contrast, conceives of learning not as a path to rulership but as a pursuit of wisdom and understanding of the world's innermost essence (Yang 1987). The *Daodejing* articulates this vision with characteristic clarity: "In the pursuit of learning, every day something is added. In the pursuit of Tao, every day something is dropped. Less and less is done, until one arrives at nonaction. When nothing is done, nothing is left undone" (为学日益，为道日损。损之又损，以至于无为，无为而无不为) (Lao Tzu 1995, chap. 48). Here, learning is not an instrument of social advancement but a contemplative path toward apprehending reality as it truly is.

Daoism therefore offers a more generative framework than Confucianism for thinking about education in an AI-shaped world. As AI disrupts education and society at large, the goal should not be to preserve existing hierarchies or reinforce entrenched power structures, but to pursue a deeper understanding of reality, as individuals and as a diverse, pluralistic community. Daoist philosophical wisdom is not merely a relic of ancient Chinese thought; in contrast to patriarchal Confucianism and to imperial and colonial Western philosophies, it carries timely and transnational resonance, conceiving of learning as the harmonious integration of self, society, and cosmos. As educators, scholars, and societies around the world grapple with the shared challenges posed by AI, each through their own cultural lenses and social conditions, traditions like Daoism may offer enduring insights that enrich contemporary thinking about learning, technology, and what it means to live well in an uncertain world.

### *The Dao Nature and the Complexity of AI in Education*

The development and transmission of knowledge have long constituted a foundational task of education (Scheffler 1978), and the epistemology of education interrogates both the essence of education and the epistemic objectives that orient it (Watson 2016). Two theoretical perspectives have shaped this field: virtue epistemology analyzes knowledge in terms of the individual knower's cognitive abilities and intellectual virtues (Battaly 2008), while social epistemology situates epistemic goods such as knowledge and understanding within social practices (Goldman 1999). Siegel (2005) contests purely truth-centered accounts, arguing that education should foster true belief in conjunction with, yet independently of, the cultivation of critical thinking and rational belief. These perspectives now confront unprecedented challenges posed by the integration of artificial intelligence into educational contexts.

Unlike earlier technological revolutions, steam-powered mechanization, electricity and telecommunications, or integrated circuits and the internet, the current wave of machine learning and large language models is profoundly embedded within the processes of teaching, learning, and knowledge production themselves (Howard 2019). AI yields tangible benefits for knowledge acquisition (Aithal and Aithal 2023), particularly in personalization, accessibility, and efficiency (Noy and Zhang 2023): adaptive learning systems tailor instruction to individual student profiles; real-time feedback accelerates the learning cycle (Ikram et al. 2026); and round-the-clock availability extends educational opportunities beyond the spatial and temporal constraints of traditional classrooms, collectively democratizing access (Pittman 2024), reducing disciplinary barriers, and enhancing overall educational productivity (Van Noorden and Perkel 2023).

Nevertheless, the realization of these advantages is contingent upon ethical implementation (Kooli 2023), encompassing robust data privacy safeguards, algorithmic bias mitigation, meaningful human oversight, and equitable infrastructure access (Kumar et al. 2025). Absent such guardrails, and given the current uneven distribution of AI, the technology risks amplifying existing educational inequities rather than ameliorating them (Stypinska 2021). The challenges are exponential and multi-dimensional, extending across three interrelated domains: knowledge itself; the epistemic processes through which knowledge is acquired; and the broader socio-environmental impacts of knowledge production. First, regarding knowledge itself, AI-generated content raises the specter of epistemic monoculture (Anderson et al. 2024) and the erosion of epistemic pluralism, alongside synthetic misinformation masquerading as authoritative truth (Agarwal et al. 2025). Second, regarding the process of knowing, reliance on cognitive offloading threatens to degrade critical thinking capacities (Gerlich 2025) and undermines academic integrity, as learners substitute algorithmic outputs for genuine intellectual engagement (Stadler et al. 2024). Third, regarding the impact of knowledge, the computational intensiveness of large-scale AI imposes high environmental costs (Wang et al. 2026), while the concentration of AI development and deployment within a small number of global technology firms reshapes geopolitical power dynamics (Pew Research Center 2025) and exacerbates structural dependencies between the Global North and South (Gabay et al. 2026).

Given this contested landscape, a re-examination and reconceptualization of the epistemic aims of education for the AI era is warranted (Müller 2025). Yet AI remains in significant respects a “black box,” resistant to full explanation even by its developers (Von Eschenbach 2021), posing a fundamental challenge to Western philosophical traditions grounded in rationalism, systematic inquiry, and justification. Moreover, the rapid pace of AI development means that epistemic goods themselves are continually being reshaped. A framework not predicated on rationalism, one oriented toward harmonizing this ongoing flux and approaching truth incrementally under conditions of uncertainty, rather than adjudicating good and bad in absolute terms, would therefore appear better suited to the current epistemic landscape of AI in education. Daoism appears particularly promising in this regard, offering conceptual resources for engaging precisely such conditions of flux and uncertainty.

Daoism is not grounded in rational analysis or logical deduction but in intuitive apprehension and holistic contemplation (Zhan and Xie 2009). The Dao, as the ultimate reality, transcends the boundaries of all concepts, language, and rational cognition and cannot be thoroughly grasped or clearly delimited by analytical thought (Hu 2005). As the *Daodejing* says, "[D]ao is beyond words, and beyond understanding. Words may be used to speak of it, but they cannot contain it" (道可道，非常道，名可名，非常名) (Lao Tzu 1995, chap. 1). Elusive and indeterminate, it resists being grasped through knowledge or analysis. "Looked at but not seen, listened to but not heard, grasped for but not held, formless, soundless, intangible, the [D]ao resists analysis and defies comprehension" (视而不见名曰夷，听之不闻名曰希，博之不得名曰微。此三者不可致诩，故混而为一) (Lao Tzu 1995, chap. 14). This inexplicability, however, is not a failure of cognition but an appropriate stance toward ultimate reality. The ultimate truth does not reside in conceptual analysis but in the integrated experience arising from the dissolution of the subject-object dichotomy (Chen 2005). This resource speaks directly to the opacity of contemporary AI. Rather than forcibly rendering everything transparent, one should acknowledge the boundaries of cognition and maintain reverence for that which cannot be fully known.

Nor can the eternal Dao even be characterized as a "being," since it is the primordial field from which all beings arise: "Dao brought forth one. One brought forth two. Two brought forth three. Three brought forth the myriad beings" (道生一，一生二，二生三，三生万物) (Lao Tzu 1995, chap. 42). The notion of "one" here refers to oneness, the ultimate ancestor of all things (Puett 2004). Rational cognition, however, depends upon distinction, definition, and inference, whereas the Dao is an undifferentiated totality, mixed and complete, dim and indistinct, prior to all discrimination. Logical and conceptual tools are therefore incapable of "capturing" the Dao; they can only gesture toward it or approximate it through metaphor. Moreover, the Dao is not a static entity but the dynamic process through which all things come into being: "moving without end or exhaustion" (周行而不殆) (Lao Tzu 1995, chap. 25). All things, caught within this ongoing flow of becoming, oscillate endlessly between complementary states, ascending and descending, light and heavy, growth and decline, *yin* and *yang* (Hu 2005). Flourishing necessarily gives way to decline, and decline and death are not endpoints but transformations into new levels of being (Kohn 2016). In this sense, the constancy of the Dao is precisely the constancy of impermanence: it abides in change itself. This dynamic, open, non-attached mode of holistic existence bears a striking resemblance to the present state of AI development and, particularly, to AI in education. Daoism therefore emerges as a source of alternative and valuable insight for exploring the complexity of AI in education.

The endeavor to comprehend AI in education in its full complexity may be understood as a contemporary form of "cultivating reality" (修真) in Daoism: not the acquisition of fixed knowledge, but an ongoing attunement to a phenomenon that perpetually exceeds complete conceptual grasp (Zhan and Xie 2009). In the Daoist tradition, "cultivating reality" does not entail becoming something other than oneself; rather, it constitutes our true reality, returning to what one originally and authentically is (Kirkland 2004). Applied to AI in education, this principle suggests that educators, learners, knowledge producers, institutions, and AI itself ought not to be alienated from their respective natures, but should each fulfill the roles proper to their authentic being.

To realize the respective nature of each entity within an AI-mediated educational world, attunement to the order of the Dao offers a viable path. "A person of Tao follows earth. Earth follows heaven. Heaven follows Tao. Tao follows nature" (人法地，地法天，天法道，道法自然) (Lao Tzu 1995, chap. 25). The Dao takes naturalness as its law, for naturalness is the very mode of being of the Dao. Here, "naturalness" does not denote the physical natural world, but rather signifies "self-so-ness," that which is so of itself, without external compulsion or deliberate contrivance. That the Dao models itself on naturalness does not imply the existence of some higher principle standing above it; rather, naturalness is the Dao's own inherent nature. As Wang Bi glosses, "The Dao does not act contrary to naturalness, and thereby attains its own nature" (道不违自然，乃得其性) (Wang Bi 1980, 65). Wuwei, non-interfering action, is the mode of praxis through which the Dao realizes this naturalness. The three challenges identified above, however, are not instances of *Wuwei* but of *Youwei (有为)*, deliberate, forced action that breaks the balance of harmony: (1) knowledge itself, in which epistemic monoculture and synthetic misinformation erode pluralism and truth; (2) the process of knowing, in which cognitive offloading degrades critical thinking and academic integrity; and (3) the impact of knowledge, in which the environmental costs of computational intensiveness and geopolitical asymmetries reinforce Global North-South dependencies. From the perspective of the Dao, all constitute departures from naturalness, thereby precluding the cultivation of reality. Dao nature, reconceived as an epistemic aim for AI in education, suggests that old currents may yet become new channels for education in the AI era.

## "Self-Cultivation (修道)" as Ongoing Practice of AI in Education

### *The Concept of "Self"*

Dao nature constitutes our true nature; the Dao is the reality that one ought to cultivate (Roth 1999). Accordingly, Daoism conceives personal transformation as occurring within a cosmos inherently structured to facilitate such change, and its core practices have centered on "self-cultivation" within a universe of subtly interconnected forces (Kirkland 2004). Crucially, this "self" is never understood as an isolated, autonomous entity endowed with inherent value superior to the external world. Rather, Daoist thought presupposes that self-cultivation is inseparable from one's relationships with others and one's embeddedness within the broader fabric of reality (Wieger 1976). This fundamentally holistic orientation, one that refuses to dichotomize self from other, distinguishes Daoist thought and practice from most other philosophical and religious traditions, whether Asian or otherwise (Zhan and Xie 2009). Moreover, Daoism has never constituted a monolithic system defined by fixed essential beliefs; rather, it represents a diverse, continually evolving cultural tradition shaped by contributors of different historical periods who brought varied ideas, values, and practices to its development (Zhao 1986). Here, “self” is flowing and heterogeneous.

In contrast to Confucian frameworks that historically excluded women from the privileged sphere of self-cultivation, Daoist conceptions of the “self” are fundamentally gender-inclusive (Jiang 2009). Classical Daoist texts, such as the *Neiye* and *Daodejing*, emerged from oral traditions whose teachings were never gender-specific (Kirkland 2004). From the Han dynasty, when organized Daoism first emerged, women maintained an active presence across its various schools. Within the *Highest Clarity (上清)* tradition, for instance, women often practiced as solitary ascetics in mixed-gender or female-only communities, with some assuming leadership of women’s congregations. Female practitioners have their own titles, including, more contemporarily, *Kundao (坤道)* (Despeux 2000), a word that joins *Kun (坤)*, symbolizing Earth, the feminine complement to *Qian (乾*, Heaven), in the *Yijing (易经)* (Huang 1991). Daoism values *Yin* more than *Yang*. Whereas Confucianism consigned women to subordinate status within patriarchal structures, *Kundao* offers a quieter yet enduring alternative (Wang 2009). This path demands the courage to reject prescribed roles; by entering the temple and renouncing conventional feminine obligations, these women claim space to construct their own identities (Wang 2009). Even in a patriarchal society, women occupied positions of genuine authority in Daoism, at times within formally institutionalized settings (Despeux 2000).

The Daoist self is also age- and social class-free. Daoism historically opened practice to all, so that self-cultivation transcended the social, gender, and generational boundaries that typically constrained individuals in other traditions (Kirkland 2004; Wang 1984). This ethos expands into a holistic worldview committed to the comprehensive transformation of self, society, and all concentric spheres of the organic universe (Chen 2005), within which the pure, empty, and spontaneous Dao-nature inheres in all conscious beings, extending even to animals, plants, and rocks (Li 2009).

If we appropriate the Daoist conception of the self, one that rejects atomistic individualism in favor of holistic embeddedness, then the very notion of being "left behind" becomes ontologically untenable. A digital divide or algorithmic bias and discrimination (Noble 2018) would constitute not merely technical or social failures, but violations of the fundamental constitution of the self, which is always already relational. Within this framework, equality of access to practice and learning is not simply an ethical injunction but an ontological given. Regardless of gender identity, age, ethnicity, geographical origin, or any other marker of social difference, all beings participate equally in the fabric of the Dao.

Moreover, this embeddedness extends even to those who occupy positions of power. The self is never merely individual; it is constituted through, and accountable to, the networks of persons, communities, and natural environments in which it is situated. Consequently, asymmetric power dynamics (Hao 2026) and the climatic and environmental costs of computational infrastructures (Kaack et al. 2022) are not externalities to be managed by detached actors, but integral dimensions of the self that demand ethical attention. By appropriating this Daoist conception of the self, the challenges confronting AI in education may be reimagined, offering alternative conceptual frameworks and renewed possibilities for navigating the human-technology relationship in education.

### *The Path of "Self-Cultivation"*

Self-cultivation is one of the central practical concerns of Daoism, a sustained effort to align oneself with the Dao and thereby transform one's body, mind, and spirit toward achieving longevity, sagehood, or transcendence (Robinet 1997). Certain practices were understood to awaken practitioners to dimensions of their own reality of which they had previously possessed only partial awareness (Wang 1984). These disciplines effect substantive personal transformation within a cosmos structured to support it, connecting the practitioner with the subtle informing structure of their own being and, to a significant extent, with their own bodily energies (Hu 2005).

Approaches to self-cultivation divide broadly into two traditions: *Waidan (外丹)*, External Alchemy, the compounding of elixirs from minerals and metals supplemented by dietary regimens, herbal medicine, and therapeutic movement practices (Gu and Dupre 2021; Han 2015), and *Neidan (内丹)*, Internal Alchemy, which accords extraordinary significance to the cultivation of interiority and psychic life (Kohn and Wang 2009). My analysis centers on *Neidan*. The *Neiye* stands as a foundational document in this regard: the earliest extant manual to systematically theorize meditation, breath control, and the physiological foundations of self-cultivation in the Chinese tradition (Huang 1991), it expounds upon the nature of the cosmos, the operations of the human mind, and an integrated suite of mental and physical disciplines aimed at self-transcendence, establishing the basic framework upon which later Daoist interiority would build (Roth 1999). Self-cultivation here entails rigorous mental disciplines centered on the refinement of "vital essence (精)," "vital energy (气)," and "numinous spirit (神)," and on psychotechnologies such as the "fasting of the mind" (心斋) and “sitting and forgetting” (坐忘), practices for the direct experiential apprehension of the Dao (Roth 1991; Wang 2011).

These self-cultivation practices are also accompanied by *Wuwei*. Practitioners cultivate emptiness, moving in unison with the formless, and the disciplined training of mind and spirit achieves a state of profound harmony (Han 2015). The transformative power of these practices derives from learning to experience and work directly with the subtle structures and energies that link personal consciousness to the broader living cosmos (Pregadio 2021). There are no cognitive shortcuts: the practitioner cannot outsource this labor to external devices but must remain experientially and cognitively engaged (Wang 2011). This insistence on sustained, first-person cultivation carries direct implications for contemporary debates on artificial intelligence in education. For educators and students alike, surrendering cognitive agency to algorithmic systems, treating them as a “thought controller”, stands in direct opposition to the Daoist project of self-cultivation, which demands the preservation of embodied, experiential engagement with reality and the Dao.

In the context of AI in education, the Daoist framework of *Neidan* compels us to reconceptualize the role of AI not as a cognitive surrogate but as an instrumental adjunct to human flourishing. Just as the Daoist practitioner cannot achieve self-cultivation through passive reception but must actively engage with the subtle energies of body and mind (Kohn and Wang 2009), so too must learners and educators resist the seduction of cognitive offloading. AI may function as a refined implement, akin to the alchemical furnace or the herbal supplement, supporting but never supplanting the labor of interior cultivation. An education grounded in Daoist principles would insist that algorithmic systems serve to amplify, rather than attenuate, the student’s direct experiential engagement with knowledge, preserving the integrity of the learning process as an embodied, effortful, and ultimately self-transformative practice. To delegate the work of thinking to machines is to sever the vital connection between consciousness and cosmos, violating the fundamental tenet that self-cultivation requires the whole person, mind, body, and spirit, to remain cognitively present and ethically accountable within the broader web of existence.

## The "Zhenren (真人)" as New Myth for AI in Education

In Daoist discourse, the term *Zhenren (真人)* first appeared extensively in the "Inner Chapters" of the *Zhuangzi*, particularly in the "Great Venerable Teacher" (庄子·大宗师) chapter (Baofeng 2025). *Zhenren* denotes one who has fully realized the Dao, an ideal personality embodying practical wisdom and morality as a whole, a complete individual harmonized with nature (Zhan and Xie 2009). As the *Zhuangzi* puts it, "To know what is done by Heaven and what is done by humans is the highest attainment. … Moreover, only when there is a *Zhenren* can there be true knowledge" (知天之所为，知人之所为者，至矣……且有真人而后有真知) (Zhuangzi 2020, chap. 6). *Zhen (真)* signifies not merely an honorific for one who has attained the ultimate aim of Daoist practice, but the very telos of that practice, "reality" itself (Chen 2005). Those who have fully integrated their being with this reality are designated *Zhenren*, "perfected persons" or "true persons," while the process of achieving such integration is termed *Xiuzhen (修真)*, or "cultivating reality" (Sun and Chen 2025), one of the deepest aims of self-cultivation. Scholars diverge on how to interpret this ideal. Roger Ames (1998) understands the *Zhenren* as an "Authentic Person" whose identity is not a static essence but a continually creative, processual mode of self-disclosure realized through cultivated spontaneity. Drawing on Robert Campany's (2009) framework for Daoist transcendents, the *Zhenren* functions as both an exemplary figure and a replicable soteriological program that practitioners approximate through specific cultural scripts and embodied techniques.

As the model of Daoist self-cultivation, *Zhenren* returns to their innate nature, transcends the artificial distinctions of social convention, and thereby dwells in equanimity, spontaneity, and cosmological unity, embodying Daoism's vision of living in harmony with universal flux, free of ego-driven striving (Sun and Chen 2025). *Zhenren* manifests a fluid, adaptive identity, responsive to the Dao's spontaneous and ever-changing rhythms and resistant to any static or essentialized form (Wieger 1976). *Zhenren* therefore presents a powerful counter-ideal to the passive, algorithmically managed learner produced by contemporary AI in education (Stadler et al. 2024).

First, *Zhenren* is a natural learner. *Zhenren* insists on first-person experiential labor, treating AI only as an intellectual prosthesis rather than a cognitive surrogate. Simultaneously, *Zhenren* follows Dao nature and rejects artificial constructions of reality, including polarized narratives surrounding AI. Rather than embracing either technological fanaticism, which portrays AI as omnipotent and a panacea for all human problems (Mohandoss 2023), or technophobia, which casts AI as a Promethean transgression, a Frankensteinian loss of control, or a transhumanist fusion of human and machine (Chaleila 2025; Hutson and Plate 2024), *Zhenren* resists frameworks that reduce the human-AI relationship to one of mastery, fear, or merger. Instead, *Zhenren* interrogates algorithmic taxonomies that sort learners and technology into ostensibly fixed categories of ability, risk, and potential, recognizing such classifications as constructed artifacts that violate the fluid, processual nature of authentic self-cultivation.

Furthermore, *Zhenren* fosters a fluid, natural knowledge partnership. Cultivating reality was never conceived as a privatized or egoistic pursuit; rather, it entailed a specialized repertoire of practices through which dedicated practitioners sought full engagement with life's most sublime and unseen forces, extending the resulting benefits to the broader community around them (Kirkland 1992). Crucially, self-cultivation entails the continuous harmonization of mind, body, and cosmos; it naturally fosters educational relations that are responsive and reciprocal rather than extractive and unidirectional. Knowledge thus circulates organically, like vital energy through a living network, with the human educator, the learner, and technology co-participating in the spontaneous generation of understanding.

Ultimately, the *Zhenren* displaces the dominant narrative of frictionless optimization with one of cultivated wholeness, reimagining learning not as data-driven efficiency but as the harmonious integration of self, society, and cosmos. The *Zhenren* functions as a replicable soteriological program rather than an elitist fantasy: this ideal remains democratically available to all. Every learner, regardless of background, gender, or class, may cultivate innate nature and relational embeddedness through disciplined practice. In doing so, the *Zhenren* myth directly resists algorithmic discrimination and asymmetric power dynamics by affirming that no student is merely a dataset to be managed, but a whole being capable of achieving equanimity, spontaneity, and cosmological unity. The educational telos shifts from producing optimized outputs to nurturing persons who remain cognitively present, ethically accountable, and holistically embedded within the broader web of existence.

## Conclusion

As artificial intelligence continues to transform educational systems around the world, the pressing question is not merely how education should adjust to AI, but what kind of persons education should cultivate within a world mediated by AI. By bringing the concepts of Dao nature, self-cultivation, and the *Zhenren* into dialogue with contemporary debates on AI, this article has shown that Daoism provides an alternative AI philosophy for education, grounded in authenticity, relationality, and harmonious participation in an ever-changing world. This study contributes to comparative philosophy of education by demonstrating that non-Western philosophical traditions can do more than diversify existing debates; they can reshape the conceptual foundations through which AI itself is understood. Careful contextualization is required, however, to avoid Orientalism in translating classical concepts into contemporary educational contexts. Future research should therefore examine how Daoist-inspired pedagogies operate across diverse educational settings and compare their insights with other non-Western traditions such as Ubuntu and Buddhist epistemology. Only by pluralizing the philosophical foundations of AI in education can we better understand what education means in an AI-shaped world.

### Positionality Statement

I write from within a bicultural scholarly training in Western philosophy of education with sustained engagement with classical Chinese texts. This location affords access to both traditions but carries risks of romanticizing Daoism, or of misreading it through Western categories, which I mitigate through the safeguards noted above (Chen 2021; Said 1995) and by treating Daoist material as a living interpretive tradition rather than a static canon.